\documentclass[sn-basic]{sn-jnl}

\usepackage{graphicx}%
\usepackage{multirow}%
\usepackage{amsmath,amssymb,amsfonts}%
\usepackage{amsthm}%
\usepackage{mathrsfs}%
\usepackage[title]{appendix}%
\usepackage{xcolor}%
\usepackage{textcomp}%
\usepackage{manyfoot}%
\usepackage{booktabs}%
\usepackage{algorithm}%
\usepackage{algorithmicx}%
\usepackage{algpseudocode}%
\usepackage{listings}%
\usepackage{tabularx}
\usepackage{orcidlink}%

\theoremstyle{thmstyleone}%
\theoremstyle{thmstyletwo}%

\theoremstyle{thmstylethree}%

\begin{document}

\title[Article Title]{Intelligence of Astronomical Optical Telescope: Present Status and Future Perspectives}


\author[1,2,3]{\fnm{Kang} \sur{Huang}\orcidlink{0000-0003-1504-9571}}\email{khuang2018@niaot.ac.cn}

\author*[1,2]{\fnm{Tianzhu} \sur{Hu}\orcidlink{0000-0002-3641-8463}}\email{tzhu@niaot.ac.cn}

\author[1,2]{\fnm{Jingyi} \sur{Cai}}\email{jycai@niaot.ac.cn}

\author[1,2,3]{\fnm{Xiushan} \sur{Pan}}\email{xspan@niaot.ac.cn}

\author[1,2]{\fnm{Yonghui} \sur{Hou}}\email{yhhou@niaot.ac.cn}

\author[1,2]{\fnm{Lingzhe} \sur{Xu}}\email{lzhxu@niaot.ac.cn}

\author[1,2]{\fnm{Huaiqing} \sur{Wang}}\email{hqwang@niaot.ac.cn}

\author*[1,2,4]{\fnm{Yong} \sur{Zhang}\orcidlink{0000-0003-2179-3698}}\email{yzh@niaot.ac.cn}

\author*[1,2]{\fnm{Xiangqun} \sur{Cui}}\email{xqcui@niaot.ac.cn}

\affil[1]{\orgdiv{Nanjing Institute of Astronomical Optics \& Technology}, \orgname{Chinese Academy of Sciences}, \orgaddress{\street{Bancang Steet}, \city{Nanjing}, \postcode{210042}, \state{Jiangsu}, \country{China}}}

\affil[2]{\orgdiv{CAS Key Laboratory of Astronomical Optics \& Technology}, \orgname{ Nanjing Institute of Astronomical Optics \& Technology}, \orgaddress{\street{Bancang Steet}, \city{Nanjing}, \postcode{210042}, \state{Jiangsu}, \country{China}}}

\affil[3]{\orgdiv{University of Chinese Academy of Sciences}, \city{Beijing}, \postcode{100049}, \country{China}}

\affil[4]{\orgdiv{National Astronomical Observatories}, \orgname{Chinese Academy of Sciences}, \orgaddress{\street{Datun Road}, \city{Beijing}, \postcode{100049}, \country{China}}}

\abstract{ With new artificial intelligence (AI) technologies and application scenarios constantly emerging, AI technology has been widely used in astronomy, and has promoted notable progress in related fields. A large number of papers have reviewed the application of AI technology in astronomy. However, relevant articles seldom mention telescope intelligence separately, and it is difficult to understand the current development status and research hotspots of telescope intelligence from these papers. This paper combines the development history of AI technology and the difficulties of critical technologies of telescopes, comprehensively introduces the development and research hotspots of telescope intelligence, conducts statistical analysis on various research directions of telescope intelligence and defines merits of the research directions. All kinds of research directions are evaluated, and the research trend of each telescope's intelligence is pointed out. Finally, according to the advantages of AI technology and the development trend of telescopes, future research hotspots of telescope intelligence are given.}

\keywords{Telescope intelligence, Mahcine learning, Site selection, Observation schedule, Datebase intelligence}



\maketitle

\section{Introduction}\label{sec1}

The potential sites for excellent astronomical observations are limited to high-altitude areas, Antarctica, and outer space, making on-site operations challenging. The use of AI to assist astronomers in harsh environments can alleviate the burden on them. AI can also enable the realization of functions that can only be achieved by complex equipment, thereby reducing equipment procurement and transportation costs and significantly alleviating the load on space telescopes. Additionally, AI can facilitate the scheduling of telescope missions and diagnose faults, enhancing imaging quality and data output.

\begin{figure}
	\centering
	\includegraphics[width=0.95\textwidth]{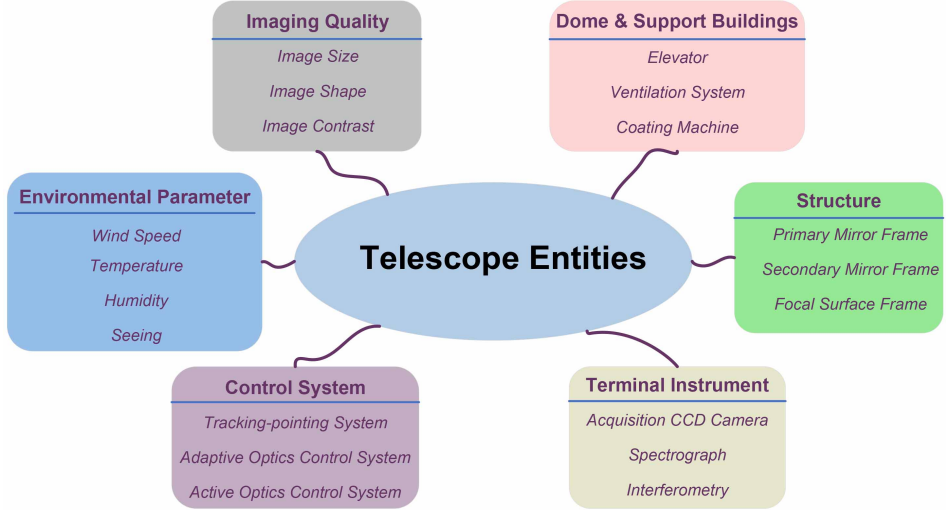}
	\caption{Classes of telescope entities}
	\label{fig:fig1}%
\end{figure}

AI technology is divided into connectionism and symbolism. Connectionism, represented by deep learning (DL), utilizes neurons for information processing, while symbolism is represented by knowledge graphs, which represent information as symbols and use rules to operate. As early as the 1990s, neural networks were applied in the observation planning of the Hubble Space Telescope (HST) \citep{johnston1992scheduling}, expert systems were used in fault diagnosis of HST energy systems \citep{bykat1990nicbes}, and statistical machine learning (ML) algorithms were widely used in the preprocessing of database data to label quasars, stars and galaxies \citep{li2008k, gao2008support, owens1996using}. With AI's evolution, its applications in telescope intelligence have broadened, encompassing selecting excellent stations, calibrating telescope optical systems, and optimizing imaging quality \citep{priyatikanto2020classification, jia2021point, gilda2022uncertainty}. In general, large ground-based astronomical telescopes are integrated optical and mechatronics devices, encompassing mechanical and drive systems, optical paths and optical systems, imaging observation, control systems, and environmental conditions. Each subsystem comprises numerous complex entities, including but not limited to the parts shown in Figure \ref{fig:fig1}. In the future, with the relentless march of cutting-edge AI technology, telescope technology will undergo significant changes.

Many articles have introduced the application of AI technology in astronomy. In 2010, \citet{ball2010data} introduced the application of traditional machine learning algorithms in astronomical big data mining, including support vector machine (SVM), artificial neural networks(ANN), K-nearest neighbor(KNN), kernel density estimation(KDE), expectation-maximization algorithm (EM), self-organizing maps (SOM), and K-means clustering algorithms. \citet{fluke2020surveying} introduced the application of the latest deep learning algorithms in the field of astronomy, and divided AI tasks into classification, regression, clustering, prediction, generation, discovery, and promotion of scientific ideas according to the type of task. \citet{meher2021deep}, and \citet{sen2022astronomical} provided a more comprehensive and detailed introduction to the application of deep learning algorithms in the field of astronomy. However, the scarcity of telescope-specific discussions in these articles makes it impossible to understand the research trends and hotspots of the current telescope intelligence. 

This paper combines the development history of AI and the difficulties of critical technologies of telescopes. It comprehensively introduces the development and research hotspots of telescope intelligence research, then conducts statistical analysis on various research directions of telescope intelligence and defines the research merits. All kinds of research are evaluated, and the research trend of each telescope intelligence direction is pointed out. Finally, according to the advantages of AI technology and the development trend of telescopes, future research hotspots of telescope intelligence are presented.

The arrangement of this article is as follows. The first part introduces the specific cases of telescope intelligence and comprehensively introduces the intelligent cases and hotspots in all stages of the telescope. The second part discusses and analyzes the current research on telescope intelligence and gives the research trends and future research hotspots. Finally, the main conclusions of the article are summarized.

\section{Telescope Intelligence}\label{sec2}

Telescope intelligence encompasses two primary domains: intelligence in manufacturing planning and intelligence in operational maintenance. The former can be further divided into astronomical observatory site selection intelligence and optical system intelligence, while the latter can be further divided into observation schedule intelligence, fault diagnosis intelligence, image quality optimization intelligence, and database intelligence.

\subsection{Observatory Site Selection}\label{subsec2}
\begin{figure}
	\centering
	\includegraphics[width=0.95\textwidth]{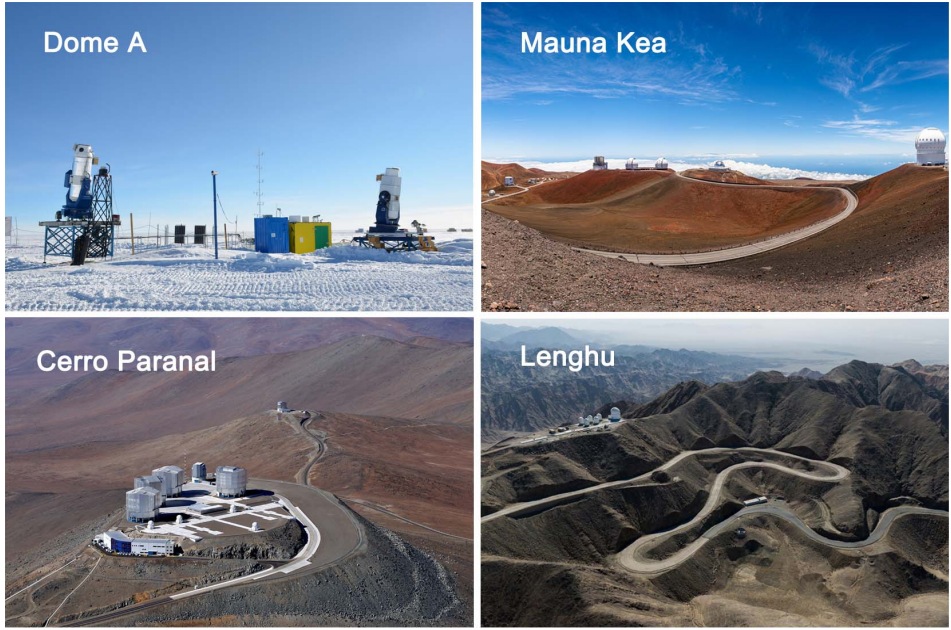}
	\caption{The environment of some sites, which are local in the high-altitude area or Antarctica}
	\label{fig:fig2}%
\end{figure}

The selection of observation sites for astronomical telescopes is crucial to maximizing their observational capabilities. Theoretical research and empirical methods for site selection have rapidly developed in recent decades, leading to the discovery of exceptional  ground-based sites such as Maunakea in Hawaii \citep{morrison1973evaluation}, Paranal and La Silla in the Chilean highlands, La Palma in Spain \citep{vernin1992optical,vernin1994optical}, Dome A in Antarctica \citep{ma2020night} and Lenghu on the Tibetan Plateau in China \citep{deng2021lenghu}. The environment of some of these sites is shown in Figure \ref{fig:fig2}.

When selecting an astronomical observatory site, it is essential to consider various observation parameters, such as the number of clear nights, atmospheric seeing, precipitable water vapor (PWV), night sky light, and meteorological parameters (such as wind speed and ground-based cloud distribution), artificial light pollution, and terrain coverage. AI methods are widely used for statistical analysis and forecasting of relevant indicators.

\subsubsection{Assessment of Site Observation Conditions}\label{subsubsec2}

In recent years, meteorological satellites, GIS technologies, and all-sky cameras have played a prominent role in assessing astronomical observation conditions at the target site \citep{aksaker2015astronomical,aksaker2020global,wang2022new}. With the use of AI technology, station cloud cover statistics can be quickly implemented.

\begin{figure}
	\centering
	\includegraphics[width=0.95\textwidth]{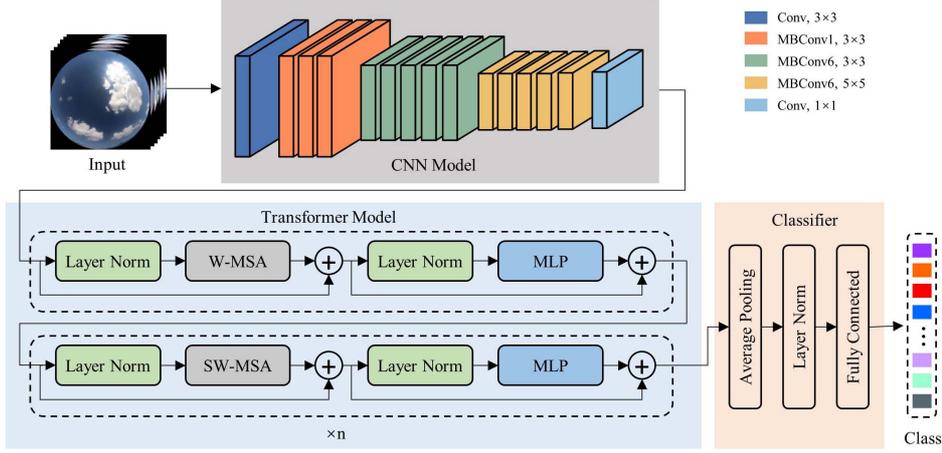}
	\caption{This method uses the output of the CNN (Convolutional Neural Network) model as input for the Transformer model to achieve cloud classification and recognition in all-sky camera imagery\citep{li2022novel}}
	\label{fig:fig3}%
\end{figure}

Conventional cloud identification, which employs multiband thresholding rules for classifying cloud areas, necessitates specific detection equipment for the corresponding bands and often yields low accuracy. SVM, principal component analysis (PCA), and Bayesian methods are utilized for single-pixel classification of satellite images, however, their performance is hindered by the absence of spatial information. \citet{francis2019cloudfcn} utilized a convolutional neural network (CNN) algorithm to amalgamate single-pixel and spatial information, realizing high-precision identification of satellite cloud conditions. \citet{mommert2020cloud} used the machine learning model based on gradient boosting, called lightGBM, and the residual neural network to classify cloud conditions from the Lowell Observatory's all-sky camera. They showed that lightGBM has superior accuracy performance. \citet{li2022novel} combined CNN and Transformer to classify and recognize cloud types, solving the drawback of the global difficulty of CNN feature extraction. The model architecture is shown in Figure \ref{fig:fig3}.

Meanwhile, employing AI techniques to classify statistical data from multiple stations can enable the prediction of PWV and sky background. \citet{molano2017low} utilized unsupervised learning to cluster meteorological parameters from various weather stations for sites across Colombia, obtaining two very low PWV astronomical sites with high probability. \citet{priyatikanto2020classification} applied random forest (RF) to classify sky brightness from different stations, enabling the monitoring of sky background brightness. Additionally, \citet{kruk2023impact} employed transfer learning and AutoML to analyze approximately two decades of Hubble Space Telescope imagery, and their work quantifies the impact of artificial satellites on astronomical observations, a factor that should be considered in the selection of  astronomical observatory sites.

\subsubsection{Site Seeing Estimate and Prediction}

\begin{figure}
	\centering
	\includegraphics[width=0.95\textwidth]{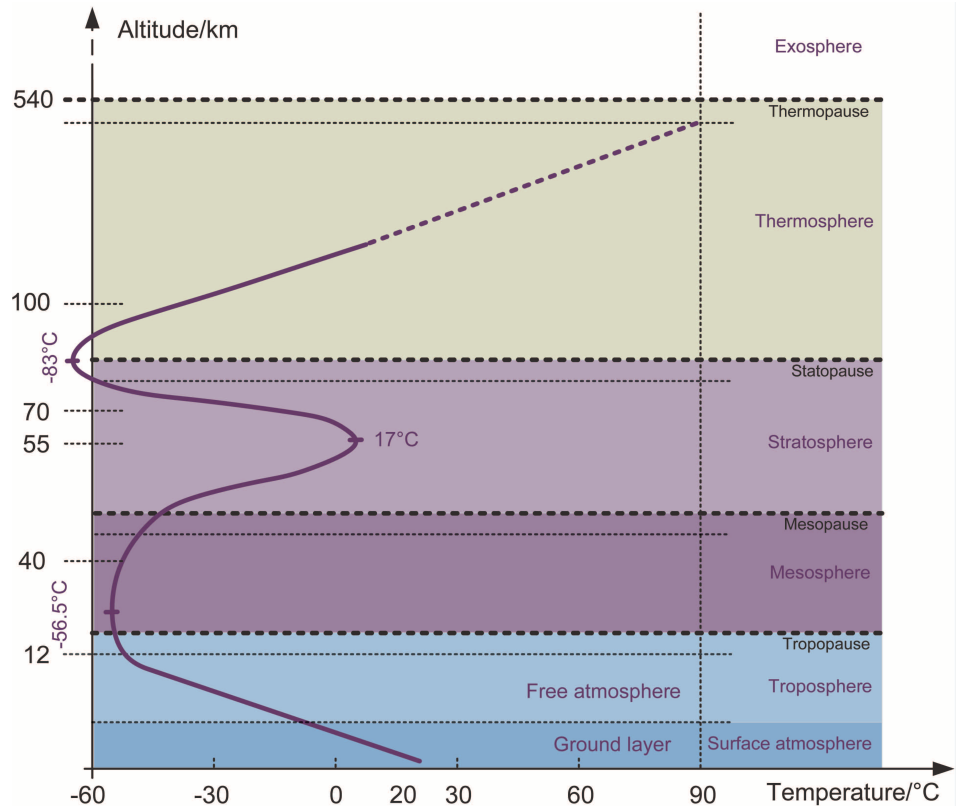}
	\caption{Schematic diagram of the structure of the Earth's atmosphere}
	\label{fig:fig4}%
\end{figure}

The structure of Earth's atmosphere, as illustrated in Figure \ref{fig:fig4}, primarily consists of the troposphere, stratosphere, mesosphere, thermosphere, and exosphere. The troposphere can be further divided by altitude into the free atmosphere and the planetary boundary layer, which has the greatest impact on atmospheric turbulence \citep{lombardi2014review,bolbasova2022atmospheric}. Monitoring and predicting atmospheric turbulence is of significant importance for enhancing the telescope observational efficiency.

As a major parameter reflecting the intensity of atmospheric turbulence, as well as the most crucial site parameter, the atmospheric seeing of observatory sites significantly affects the image quality of optical telescopes. Certain models estimate atmospheric seeing by correlating atmospheric parameters with the integrated parameters of optical turbulence (astronomical optical parameters), such as the Dewan model \citep{dewan1993model}, the Coulman-Vernin model \citep{coulman1988outer}, and the AXP model proposed by Trinquet \& Vernin \citep{trinquet2006model}. However, these models, being based on statistical data from multiple stations, are less effective for specific site analysis. AI techniques are used to construct relationships between atmospheric parameters and astronomy optical parameters for a given station. 

$C_N^2$, a key parameter reflecting the change of optical turbulence intensity, is also an important parameter for deriving atmospheric seeing through the analytical model. In a pioneering effort, \citet{wang2016using} from the Mauna Loa Observatory first used ANNs on atmospheric temperature, pressure, and relative humidity to estimate the structure constant of the refractive index $C_N^2$. \citet{jellen2020machine} used the RF method to predict the $C_N^2$ of the near-surface atmosphere and studied the contribution of environmental parameters to optical turbulence. \citet{su2020adaptive} operated an optimized backpropagation (BP) network to experiment with data obtained from Chinese Antarctic scientific research, showing that $C_N^2$ forecasting results based on this method had a reliable correlation. \citet{vorontsov2020atmospheric} processed the short-exposure laser beam intensity scintillation patterns based on deep neural networks (DNNs) to predict $C_N^2$, achieving superior measurement accuracy and higher temporal resolution. \citet{bi2022optical} used a GA-BP neural network to train and predict the meteorological parameters collected by the instrument, which can deduce the relevant astronomical optical parameters. \citet{grose2023forecasting} employed a turbulence prediction method based on recurrent neural networks (RNN), utilizing prior environmental parameters to forecast turbulence parameters for the next 3 hours. 

AI techniques have also been used for the direct forecasting of seeing. \citet{kornilov2016forecasting} analyzed the atmospheric optical turbulence data above Mount Shatdzhatmaz and predicted the short-time seeing. \citet{milli2020turbulence} constructed a seeing prediction strategy based on the RF for the Paranal Observatory to provide a reference for optimizing telescope observation efficiency. \citet{giordano2020statistical} used the atmospheric parameter database of the Calern Observatory as an input for statistical learning to make predictions of turbulence conditions while illustrating the importance of the in situ parameter characteristics of the station \citep{giordano2021contribution}. The Maunakea Observatory has used its observation and forecast data to build a machine-learning seeing prediction model that can make predictions for the following five nights \citep{lyman2020forecasting,cherubini2022forecasting}. \citet{turchi2022optical} used the RF method to make a short-scale time (1-2 hours) forecast of atmospheric turbulence and seeing above the Very Large Telescope (VLT). \citet{hou2023machine} employed wind speed and temperature gradient acquired from Antarctic Dome A as inputs and predicted the seeing based on long short term memory (LSTM). \citet{masciadri2023optical} introduced a method for short-term (1-2 hours) prediction of astroclimate parameters, including seeing, airmass, coherence time, and ground-layer fraction, demonstrating its effectiveness at the VLT. \citet{ni2022data} used data from the Large Sky Area Multi-Objective Fiber Spectroscopic Telescope (LAMOST), the development and research of seeing prediction models using main AI techniques, including statistical models ARIMA and Prophet, machine learning methods Multilayer Perceptron (MLP) and XGBoost, and deep learning methods LSTM, Gate Recurrent Unit (GRU), and Transformer. The method, input parameters, output parameters and accuracy of the above methods are shown in Table \ref{tab1}.

\begin{sidewaystable}
	\caption{Method, input parameters, output parameters and accuracy of different method}\label{tab1}
	
	\begin{tabularx}{\textwidth}{@{\extracolsep\fill}XXXX}
		\toprule%
		Method & Input Parameters & Output Parameters & Accuracy \\
		\midrule
		ANN\citep{wang2016using} & temperature, relative humidity, pressure, potential temperature gradient, wind shear   &  $C_N^2$   &    $R^2=0.87$, weekly  \\
		RF\citep{jellen2020machine} & dew point temperature, pressure, wind, relative humidity, et al.  &  $C_N^2$ & $MSE$=0.09  \\
		Optimized BP\citep{su2020adaptive}  &  pressure, temperature, relative humidity, wind speed, snow face temperature  & $C_N^2$  &  $R_{xy}=0.9323$ \& $RMSE$=0.2367  \\
		DNN\citep{bi2022optical}  &  simulated $C_N^2$   &  $C_N^2$   &  normalized $RMSE$=0.072 \& $std$=0.06   \\
		GA-BP\citep{vorontsov2020atmospheric} & height, pressure, temperature, wind speed, wind shear \& temperature gradient    &   $\log{(C_N^2)}$    & $RMSE<1.4$ \\
		RF \& MLP\citep{milli2020turbulence} & seeing, atmospheric parameters (pressure, temperature, wind, humidity, PWV)  & seeing  & $RMSE=0.27$, 2 hours    \\
		RF\citep{giordano2020statistical} & ground parameters (wind, temperature, relative humidity, pressure) \& integrated coherence parameters   & seeing   & Pearson correlation coefficient = 0.8  \\
		k-means\citep{cherubini2022forecasting} & free seeing \& its std, wind velocity, shear from GFS  & seeing of total \& free atmospheric of next 5 days   & $RMSE<0.25$  \\
		RF\citep{turchi2022optical}  & seeing, wavefront coherence time, isoplanatic angle, ground layer fraction and atmospheric parameters (temperature, relative humidity, wind speed \& direction)  & seeing   & $RMSE=0.24$, 1h $RMSE=0.32$, 2h  \\
		LSTM \& GPR\citep{hou2023machine} & wind speed \& temperature gradient  &  seeing & $RMSE=0.14$, 10min  \\
		\botrule
	\end{tabularx}
\end{sidewaystable}

\subsection{Intelligence of Optical Systems}

The telescope's optical system is the most crucial part of the telescope. Optical system misalignment directly affects the imaging quality, resulting in deformation of the star's shape and the enlargement of the star's size. Traditional optical system calibration relies on laser interferometer and wavefront detection equipment, which are challenging to operate in harsh environments. AI technology can replace or simplify equipment operations to achieve optical path and mirror surface calibration.

\subsubsection{Optical Path Calibration}

Large-aperture and wide-field survey telescopes often have a primary mirror with a fast focal ratio, making the secondary mirror more sensitive and requiring higher calibration accuracy \citep{wu2022machine}. Using lasers interferometer and other equipment for manual adjustment, the mirror tilt accuracy can reach ten arc seconds, and the eccentricity accuracy can reach 0.1 mm \citep{li2015alignment}. Many computer-aided alignment methods have been developed to achieve higher-precision calibration, including vector aberration theory proposed for the optical path alignment of LSST (Large Synoptic Survey Telescop) and JWST (James Webb Space Telescope) \citep{thompson2008misalignment}. With mature wavefront detection technology, the method of wavefront detection and inverse calculation of misalignment error based on Zernike polynomial decomposition has played an enormous role in the collimation and adjustment process of the telescope.

The method mentioned above requires the wavefront sensor device and the relationship between the telescope's aberration and the adjustment error and meets the challenge of adjusting multiple fields of view simultaneously. AI technology can construct the relationship between the adjustment error and the detection parameters. \citet{wu2022machine} used ANN to construct the relationship between the star image obtained by the scientific camera and the adjustment error of the telescope to realize optical path calibration. \citet{jia2021point} proposed a CNN-based algorithm to fit the relationship between the point spread function (PSF) and four degrees of freedom of the secondary mirror, which can be used to align the secondary mirror of wide-field survey telescopes. The Rubin Observatory uses CNN with a self-attention mechanism to realize the correspondence between the degrees of freedom of the primary mirror, secondary mirror, and focal plane and the final imaging of the science camera and realizes the active adjustment of the attitude of the LSST telescope \citep{yin2021active}.

In addition to being applied to calibrating the whole optical path of the telescope, AI technology can also be applied to calibrating a specific device of the telescope. LAMOST is a large-field survey telescope with a focal surface having 4,000 optical fibers. Before each observation, the optical fiber needs to be moved to the corresponding position. The positioning accuracy of the optical fiber is closely related to the initial angle of the optical fiber head. CNN is used to classify the camera's pixels, including the optical fiber head, to achieve the optical fiber's contour extraction and the optical fiber's initial angle \cite{zhou2021lamost}.

\subsubsection{Mirror Surface Calibration}

Active optics technology is the primary means to realize the surface shape calibration of large-aperture telescopes. The active optical system can be divided into two steps. The first part is wavefront reconstruction, including phase retrieval, the phase diversity method, and the wavefront sensor-based method. The second part obtains the calibration voltage according to the corresponding relationship between the wavefront and the calibration voltage. The calibration voltage is imported into the force actuator to realize the adjustment of the surface shape \citep{su2004active}.

\begin{figure}
	\centering
	\includegraphics[width=0.95\textwidth]{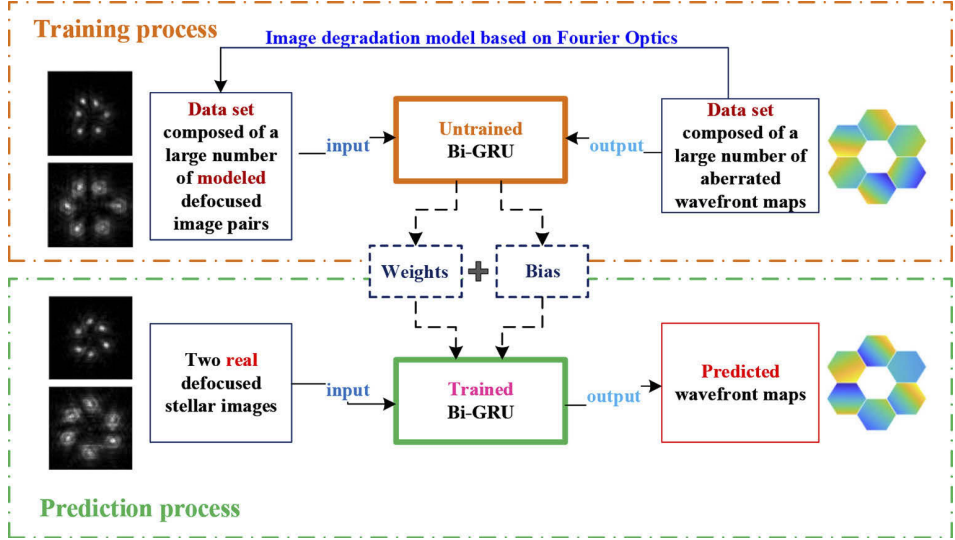}
	\caption{In focus and defocus star image and aberrated wavefront maps are used to traine the Bi-GRU network, and trained network is used to predict wavefront maps \citep{wang2021deep}}
	\label{fig:fig5}%
\end{figure}

AI technology can be applied to active optical technology to calibrate the surface shape. The DNN can be used to directly construct the relationship between the point map obtained by the Shack-Hartmann wavefront sensor and the calibration voltage to improve the calibration efficiency \citep{li2020deep}. Bi-GRU can be used to obtain the corresponding relationship between the defocus star images and the wavefront \citep{wang2021deep}. The sketch map of the co-phasing approach using the Bi-GRU network is shown in Figure \ref{fig:fig5}. SVM can be utilized to overcome the shortcomings of curvature sensing that are easily affected by atmospheric dis-turbances and improve the calibration ability of curvature sensing \citep{cao2020extending}.

Segmented mirror telescope surface calibration mainly includes piston error and tip/tilt error, among which tip/tilt error detection is relatively mature, while piston error detection is still challenging, and $2\pi$ error is prone to occur. ANN can be used to build the relationship between the piston error and the amplitude of the modulation transfer function (MTF) sidelobes to detect the piston error \citep{yue2021piston}. CNN can be used to distinguish the range of the piston error and improve the sensitivity of the Shack-Hartmann wavefront sensor to the $2\pi$ error \citep{li2019large}. \citet{wang2021multichannel} proposed a CNN-based multichannel left-subtract-right feature vector piston error detection method, which can improve the detection range of the piston error to $-139\lambda-139\lambda$.

\subsection{Intelligent scheduling}

The telescope executes observation tasks according to the observation schedule. A long-term scheduler refers to allocating observation time for observation tasks in the next few months, selecting valuable observation proposals from numerous observation proposals, and then conducting observations according to the proposal's priority, observation time, observation goals, and the constraints of the telescope itself. At the same time, it is necessary to formulate short-term planning and adjust the observation plan for new targets or transient phenomena.

Observation planning requires the calculation of many observation tasks to find the best observation plan. Manual planning is not sufficient for long-term task planning. Research in this area has been ongoing since the mid-1950s, ranging from simple heuristics to more complex genetic algorithms or neural networks. AI techniques are widely used in observation planning tasks. \citet{granzer2004makes} introduced traditional observation planning methods, including queue planning, critical path planning, optimal planning, and allocation planning. \citet{colome2012research} further comprehensively introduced observation planning techniques over the past 50 years, mainly based on genetic algorithm, genetic algorithm, ant colony optimization algorithm, multiobjective evolutionary algorithm and other new methods.

In the 1990s, the Hubble Space Telescope utilized an ANN-based SPIKE system \citep{johnston1992scheduling,johnston1994spike} to generate observation plans, and extended its use to the VLT and Subaru telescopes. The non-dominated sorting genetic algorithm (NSGA-II) was employed in DSAN, RTS2, EChO project \citep{garcia2015artificial}, and the 3.5-meter Zeiss telescope \citep{garcia2017efficient}. It is suitable for long-term planning tasks and can be used in conjunction with constraint-based methods. The generalized differential evolution 3 (GDE3) algorithm is more efficient than NSGA-II, and it was combined with SPIKE to provide observation planning for JWST \citep{adler2014planning}. In addition, it is also used in the DSAN project. The SWO optimizer based on the greedy algorithm is used in SOFIA \citep{frank2006sofia}, Mars Rover, and THEMIS projects. Reinforcement learning is used for the planning of LSST telescopes and the ordering of sky areas observed by optical telescopes, improving the probability of optical telescopes discovering transient astronomical phenomena such as gravitational waves, gamma-ray bursts, and kilonovae \citep{astudillo2023reinforcement,naghib2019framework}.

\subsection{Fault Diagnosis}

Real-time fault monitoring and efficient diagnosis can prevent observation time waste and ensure high-quality imaging. In addition, equipment faults in telescopes can lead to significant economic losses. Traditional telescope fault diagnosis involves installing numerous sensors to monitor telescope parameters, such as voltage and meteorological conditions, and setting thresholds based on experience. When these parameters exceed the threshold, the alarm system will issue a warning and identify the location and cause of the fault.

The application of AI technology in fault diagnosis has a long history. Since the 1980s, fault diagnosis systems based on expert knowledge have been widely used in fields such as aerospace, automotive fault diagnosis, and telescopes. For instance, \citet{dunham1987knowledge} applied a knowledge-based diagnostic system to achieve fault diagnosis of the pointing and tracking system of the Hubble Space Telescope, while \citet{bykat1990nicbes} used an expert system to diagnose faults in the energy system of the same telescope. The fault diagnosis system based on expert knowledge has the advantages of strong logic and intuitive knowledge representation, and is still widely used. \citet{yun2018reliability} used the knowledge tree to implement the intelligence of the main-axis control system of the Antarctic Sky Survey Telescope AST3. \citet{tang2023fault} proposed a method for the rapid localization of faults in LAMOST's fiber positioner fault causes based on LSTM.

\begin{figure}
	\centering
	\includegraphics[width=0.95\textwidth]{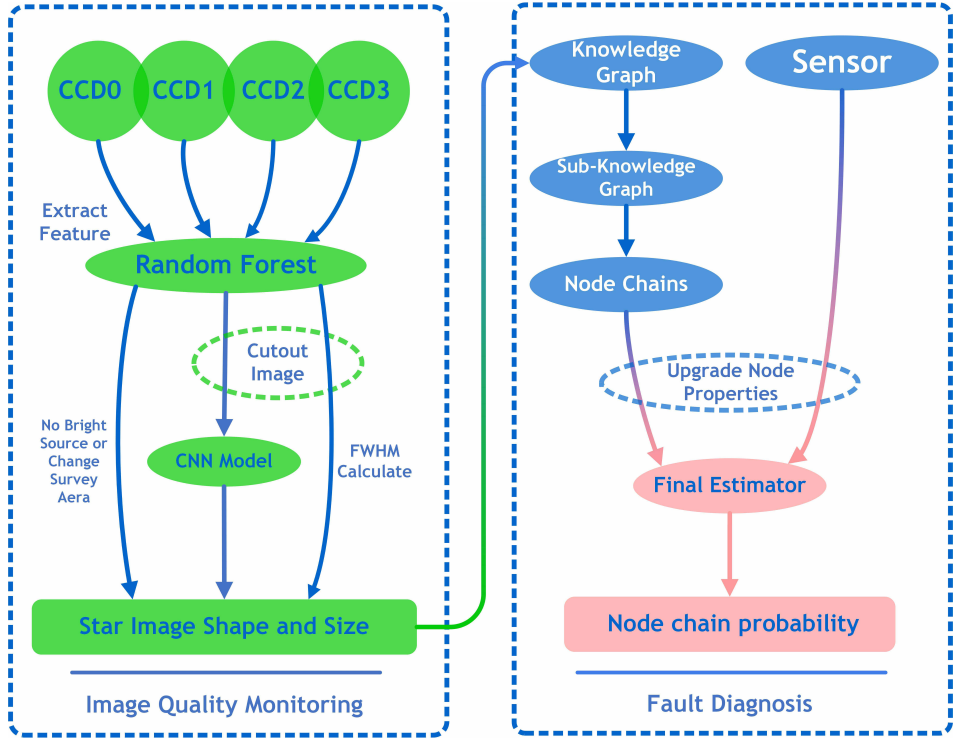}
	\caption{In the framework of the telescope maintenance support system, the left part realizes the imaging quality monitoring, and the right part realizes the fault diagnosis\citep{hu2023intelligent}}
	\label{fig:fig6}%
\end{figure}

In the 1980s, fault diagnosis systems based on ANNs gained popularity with the rise of neural networks. The fault diagnosis problem can be viewed as a classification problem, where the correspondence between sensor features and faults can be established. In recent years, deep learning has advanced the use of neural networks for deep fault feature extraction. For instance, \citet{teimoorinia2020assessment} combined SOM and CNN to classify star image shapes for telescope imaging, enabling timely detection of poor-quality telescope imaging. Similarly, \citet{hu2021telescope} used CNNs to establish a relationship between telescope failure and telescope images, enabling the initial diagnosis of telescope faults. Recently, we also have proposed a methodology for real-time monitoring and diagnosis of the imaging quality of astronomical telescopes \citep{hu2023intelligent}, incorporating AI technologies such as CNN and knowledge graphs, and validated it using observational data from LAMOST. This holds profound implications for the future of monitoring and diagnosing imaging quality in next-generation large telescopes. The framework of this approach is shown in Figure \ref{fig:fig6}. 

Neural networks are also being used for fault prediction by establishing a relationship between parameters before a fault occurs and the probability of the fault occurring. However, fault prediction on telescopes is still in its early stages due to the lack of training data.

\subsection{Optimization of imaging quality}

The imaging quality of an astronomical telescope can be quantified by the full width at half-maximum (FWHM) of the light intensity distribution. A smaller FWHM indicates better imaging quality. The measured values of imaging quality (${IQ}_{Measured}$) are mainly affected by defects in the optical system, turbulence introduced by the dome, and atmospheric turbulence, expressed with corresponding indicators (${IQ}_{Optics}$, ${IQ}_{Dome}$, and ${IQ}_{Atmosphere}$, respectively). These indicators are assumed to conform to Kolmogorov's Law, and the following relationship exists:

\begin{equation}
	{IQ}_{Measured}^{5/3}={IQ}_{Optics}^{5/3}+{IQ}_{Dome}^{5/3}+{IQ}_{Atmosphere}^{5/3}
\end{equation}

Assuming the observatory site is confirmed and the defects in the telescope's optical system cannot be further improved, optimizing the imaging quality can be achieved by addressing atmospheric turbulence and dome seeing. This can be done through the use of adaptive optics technology to calibrate atmospheric turbulence and improve the dome design and ventilation system to enhance dome seeing.

\subsubsection{Dome seeing}

The primary factors that affect dome seeing are the local temperature differences inside the dome, especially near the primary and secondary mirrors, and turbulent flow caused by temperature differences inside and outside the dome. Therefore, controlling the temperature difference is crucial in reducing the deterioration of imaging quality caused by dome seeing.   

In general, the air temperature inside the observatory dome is higher than the ambient temperature at night. The initial solution was to ventilate the dome for a few hours before nighttime observation to bring the temperature inside the dome in line with the ambient temperature. However, due to the high-temperature inertia of the primary and secondary mirrors, more than a few hours of ventilation is needed to bring the temperature down to the same level as the ambient temperature. As a result, a local turbulence layer, known as mirror seeing, can form near the mirror with a higher temperature, leading to a deterioration of imaging quality.

To mitigate this issue, the general practice is to use temperature control means, such as air conditioning, during the daytime to adjust the temperature inside the dome, primarily the temperature of the primary and secondary mirrors, to be consistent with the ambient temperature during nighttime observations. Accurate prediction of the ambient temperature during nighttime observations is crucial for temperature control.

\citet{murtagh1993nowcasting} utilized meteorological data recorded by the European Southern Observatory in La Silla and Paranal to predict temperature 24 hours later using the KNN method. They used the current, 24-hour-ahead, and 48-hour-ahead temperatures, as well as the current and 24-hour-ahead air pressure to achieve an accuracy of 85.1\% within an error range of 2\textcelsius, with a 70\% reliability of predicting good seeing. \citet{aussem1995dynamical} investigated the accuracy and adaptability of dynamic recurrent neural networks and KNN for time series prediction using the same data. They found that many interruptions in the recording sequence and insufficient data were the main limiting factors of these two methods. Additionally, they proved that in the case of fuzzy coding of seeing (dividing the seeing degree into Good, Moderate, and Bad), the forecasting accuracy of dynamic recurrent neural networks (RNNs) outperforms that of KNN. \citet{buffa1997temperature} studied the problem of forecasting observatory site temperatures and proved that the nonlinear autoregressive neural network model is more competitive than the traditional linear filtering algorithm.

\begin{figure}
	\centering
	\includegraphics[width=0.95\textwidth]{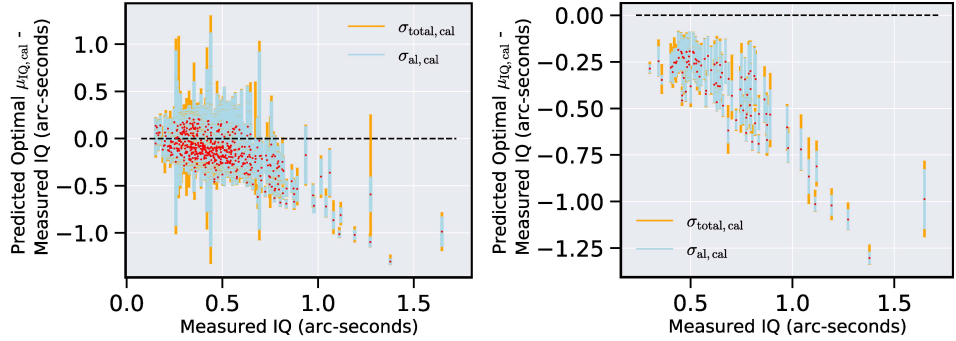}
	\caption{The baseline configuration is all-open. After restricting ourselves to a subset of ID 'togglings', figure plot the improvement over the measured MPIQ values \citep{gilda2022uncertainty}}
	\label{fig:fig7}%
\end{figure}

In addition, \citet{gilda2022uncertainty} developed a method for predicting the probability distribution function of observed IQ based on environmental conditions and observatory operating parameters using Canada France Hawaii Telescope (CFHT) data and mixture density network (MDN). They then combined this approach with robust variational autoencoder (RVAE) to forecast the optimal configuration of 12 vents to reduce the time required to reach a fixed signal-to-noise ratio (SNR) for observations. This approach has the potential to increase scientific output and improve the efficiency of astronomical observations. Figure \ref{fig:fig7} presents results on the improvement in MPIQ predicted by MDN given the (hypothetically) optimal vent configurations selected from the restricted set of ID configurations selected by the RVAE.

\subsubsection{Adaptive optics}

Adaptive optics (AO) is a crucial tool for improving the imaging quality of ground-based optical telescopes by ameliorating atmospheric turbulence. Since its first implementation in the telescopes of the European Southern Observatory 30 years ago, it has been widely used for imaging quality optimization. \citet{guo2022adaptive} introduced machine learning methods in adaptive optics, including improving the performance of wavefront sensors, building WFSless AO systems, and developing wavefront prediction techniques.

Improving the performance of traditional wavefront sensors is similar to the content of active optics, which includes enhancing the anti-noise of wavefront detection \citep{li2018centroid,guo2006wavefront} and constructing the relationship between wavefront detection equipment imaging and wavefront \citep{suarez2018improving,dubose2020intensity}. In addition to being implemented in conventional adaptive optics systems, AI techniques have also been utilized to overcome the sensitivity of multiobjective adaptive optics systems to atmospheric contour changes \citep{osborn2014first}.

\begin{figure}
	\centering
	\includegraphics[width=0.95\textwidth]{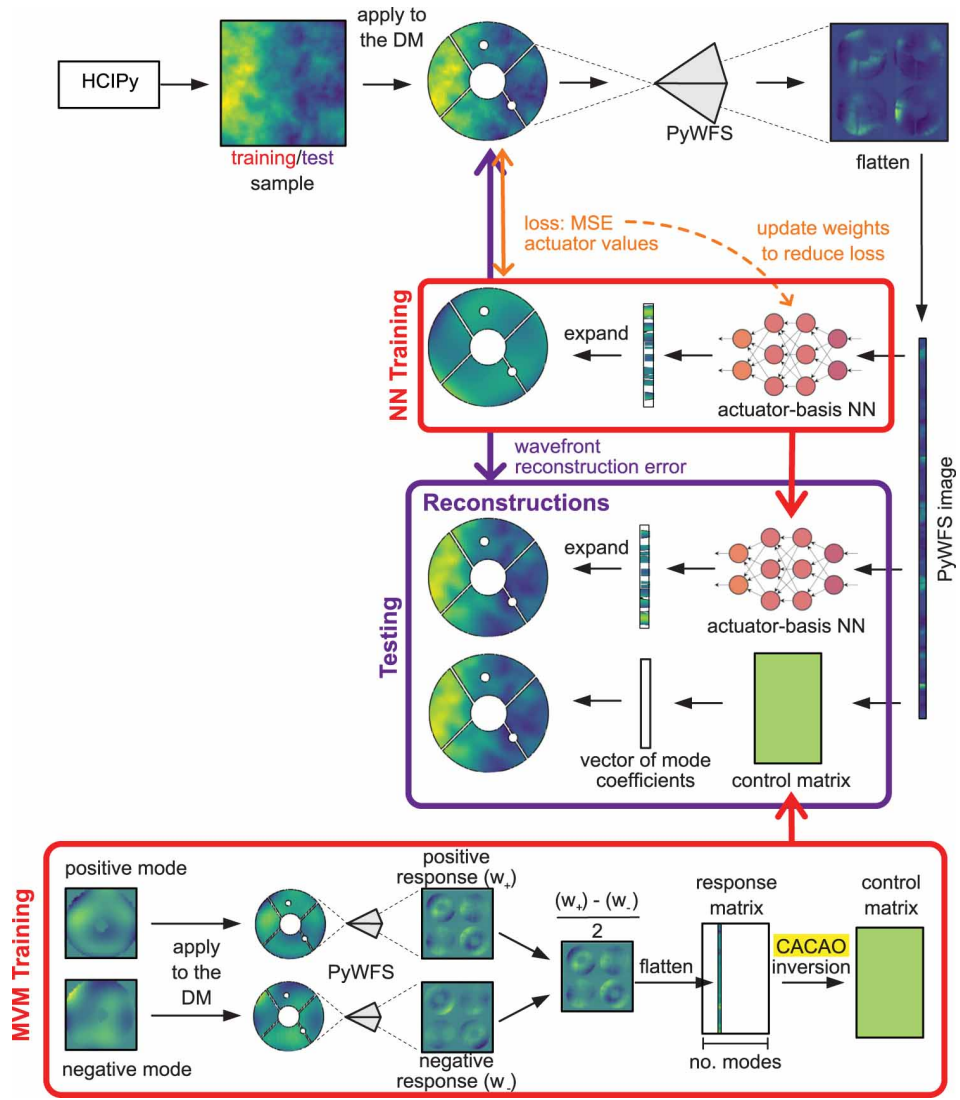}
	\caption{ Visualization of the training and testing process on the turbulence-like data set mentioned in related paper \citep{wong2023nonlinear}}
	\label{fig:fig8}%
\end{figure}

WFSless systems do not use conventional wavefront sensing devices to construct wavefronts. \citet{kendrick1994phase} used defocusing images to generate wavefronts based on neural networks. \citet{wong2023nonlinear} ultilized bottleneck networks for more precise wavefront reconstruction, which has been confirmed to have enhanced performance by data from the Adaptive Optics (AO) system of the Subaru Telescope, some training and testing procedures of the proposed method are shown in Figure \ref{fig:fig8}. With the wide application of deep learning, CNN is also employed for WFSless systems for wavefront detection \citep{swanson2018wavefront,guo2019improved,ma2019numerical,wu2020sub}.

The calibration frequency of adaptive optics is much higher than that of active optics technology, and the reconstruction and feedback of the wavefront make the wavefront calibrated by the optical calibration equipment have an unavoidable time delay compared with the actual wavefront. Improving the speed of wavefront reconstruction or predicting future wavefront can solve this problem. \citet{montera1997prediction} compared the prediction accuracy of the linear minimum mean square error method and the neural network algorithm, indicating that the prediction performance of the latter is better. The recurrent neural network LSTM \citep{liu2020wavefront} and the Bayesian regularization-based neural network architectures \citep{sun2017bayesian} are also implemented for wavefront prediction.

In addition, AI technology is also used to expand the ultra-high-resolution imaging of light sources. This includes using CNNs to construct encoding and decoding layers, where the encoding layers extract image features, and the decoding layers output corrected images and multi-frame information to improve solar imaging resolution \citep{ramos2018real}. Generative adversarial network (GAN) is also utilized to generate high-resolution solar magnetic field pictures \citep{kim2019solar,rahman2020super}. However, as this research content does not belong to traditional telescope technology, we will not further introduce it as a telescope-related intelligent technology.

\subsection{Database Intelligence}

The astronomical database system serves as a service platform for data storage and sharing, allowing astronomers and other relevant users worldwide to share, obtain, and mine valuable information from astronomical database lists \citep{ribeiro2013survey}. To enhance the scientificity and richness of the data, it is crucial to establish an intelligent astronomical database. AI technology is well-suited for database data fusion and classification due to its ability to automatically extract features \citep{fluke2020surveying}.

\subsubsection{Database Data Fusion}

The Astronomical Database has been accumulating data since the 1980s and comprises various sub-databases, such as the astronomical star list database, large-field multicolour sky survey database, asteroid database, and astronomical literature database. The cross-fusion of databases is a significant trend in current astronomy. Table \ref{tab2} shows the classification of the database.

\begin{table}[thp]
	\caption{Classification of the database and representative catalogue}
	\centering
	\label{tab2}
	\setlength{\tabcolsep}{2mm}{
		\begin{tabular}{@{}ccc@{}}
			\hline
			Catalog Database & Volume & Representative Catalogue \\
			\hline%
			\multirow{2}{*}{I.Astrometric Data} &   \multirow{2}{*}{1136}   &  AGK3 Catalogue(I/61B)\\
			&   & UCAC3 Catalogue(I/315)  \\
			\multirow{2}{*}{II.Photometric Data} &   \multirow{2}{*}{747}   &  General Catalog of Variable Stars, 4th Ed(II/139B)\\
			&   & BATC--DR1(II/262)  \\
			\multirow{2}{*}{III.Spectroscopic Data} &   \multirow{2}{*}{291}   &  Catalogue of Stellar Spectral Classifications(III/233B)\\
			&   & Spectral Library of Galaxies, Clusters and Stars(III/219)  \\
			\multirow{2}{*}{IV.Cross-Identifications} &   \multirow{2}{*}{19}   &  SAO-HD-GC-DM Cross Index(IV/12)\\
			&   & HD-DM-GC-HR-HIP-Bayer-Flamsteed Cross Index(IV/27A)  \\
			\multirow{2}{*}{V. Combined Data} &   \multirow{2}{*}{554}   &  The SDSS Photometric Catalogue, Release 12(V/147)\\
			&   & LAMOST DR5 catalogs(V/164)  \\
			\multirow{2}{*}{VI.Miscellaneous} &   \multirow{2}{*}{379}   &  Atomic Spectral Line List(VI/69)\\
			&   & Plate Centers of POSS-II (VI/114)  \\
			\multirow{2}{*}{VII.Non-stellar Objects} &   \multirow{2}{*}{1136}   &  AGK3 Catalogue(I/61B)\\
			&   & UCAC3 Catalogue(I/315)  \\
			\multirow{2}{*}{I.Astrometric Data} &   \multirow{2}{*}{292}   &  NGC 2000.0(VII/118)\\
			&   & SDSS-DR5 quasar catalog(VII/252)  \\
			\multirow{2}{*}{VIII.Radio and Far-IR Data} &   \multirow{2}{*}{99}   &  The 3C and 3CR Catalogues(VIII/1A)\\
			&   & Miyun 232MHz survey(VIII/44)  \\
			IX.High-Energy Data &   47  &  Wisconsin soft X-ray diffuse background all-sky Survey(IX1))\\
			\hline
	\end{tabular}}
\end{table}

The Bayesian-based method is widely used in astronomical catalog cross-matching and image fusion \citep{yu2019astronomical}. \citet{budavari2008probabilistic} developed a unified framework, grounded in Bayesian principles, for object matching, which includes both spatial information and physical properties. Additionally, \citet{medan2021bayesian} presented a Bayesian method to cross-match 5,827,988 high proper-motion Gaia sources with various photometric surveys. Furthermore, Bayesian methods are employed to determine the probability of whether the data represent objects or the background in image fusion \citep{jalobeanu2008multi,petremand2012optimal}.

To enhance the connection between various data sets, it is necessary to strengthen the application of AI in database retrieval and outlier retrieval. As an example, \citet{du2016efficient} developed a method based on the bipartite ranking model and bagging techniques, which can systematically search for specific rare spectra in the SDSS spectral data set with high accuracy and low time consumption. Similarly, \citet{wang2022unsupervised} proposed an unsupervised hash learning-based rare spectral automatic approximate nearest neighbor search method, which searches for rare celestial bodies based on spectral data and retrieved rare O-type stars and their subclasses from the LAMOST database. 

Retrieving outliers in the database can improve reliability and realize the fusion of astronomical databases. \citet{rebbapragada2009finding} combined PCAD (Periodic Curve Anomaly Detection) with the K-means clustering algorithm to separate anomalous objects from known object categories. However, this method has poor scalability and may lose possible outliers in massive data sets and high-dimensional space. Therefore, \citet{nun2014supervised} proposed a new method based on RF to automatically discover unknown abnormal objects in large astronomical catalogs and followed up on the outliers for more in-depth analysis by cross-matching with all publicly available catalogs. Moreover, multiple data sources may lead to the loss of a certain amount of association information in different data sets. To address this issue, \citet{ma2021outlier} proposed an outlier detection technique combining new density parameters of KNN and RNN to mine relevant outlier information from multisource mega-data sets.

\subsubsection{Database Data Labeling}

Astronomical databases store an enormous amount of features of astronomical data, which gives rise to dimensional problems that are difficult to analyze. Therefore, the utilization of AI to analyze and annotate celestial information parameters in the astronomical database is of great significance for further astronomical research. 

\paragraph{Automatic Data Classification}

The exponential growth of astronomical data has made automatic classification of data generated from large-scale surveys crucial. The automatic classification of celestial data includes the classification of quasars, galaxies, and stars based on various features, such as their spectra, luminosity, and other celestial information.

\citet{banerji2010galaxy} performed morphological classification of galaxy samples from SDSS DR6 based on ANNs and compared it with the Galaxy Zoo. \citet{ball2010data} used spectral data from the third SDSS data release to train KdTree to provide reliable classification research for all 143 million non-repetitive photometric objects. \citet{zhang2009automated} combined KNN and RF to separate quasars and stars. \citet{aguerri2011revisiting} applied SVM to automatically classify approximately 700,000 galaxy samples from SDSS DR7 and provided the probability that each sample belongs to a certain category.

Additionally, unsupervised techniques have been used to classify astronomical data. \citet{mei2019unsupervised} adopted a three-dimensional convolutional autoencoder (3D-CAE) to implement the unsupervised spatial spectral feature classification strategy. \citet{fraix2021unsupervised} used the unsupervised clustering Fisher-EM algorithm to classify galaxies and quasars with spectral redshift less than 248.0 in the SDSS database.

Moreover, deep learning has been used to analyze high-dimensional spectral data to classify astronomical objects. \citet{khalifa2017deep} proposed a deep CNN structure for galaxy classification with high testing accuracy. \citet{becker2020scalable} introduced a scalable end-to-end recurrent neural network (RNN) scheme for variable star classification, which can be extended to large datasets. \citet{hinners2018machine} discussed the effectiveness of LSTM and RNN deep learning in stellar curve classification. \citet{awang2020classification} used transfer learning to classify planetary nebula (PNe) in the HASH DB and Pan STARRS databases. \citet{barchi2020machine} combined accurate visual classifications from the Galaxy Zoo project with machine learning and deep learning methodologies to improve galaxy classification in large datasets.

\paragraph{ Preselect Quasar Candidates}

Quasars are a type of active galactic nucleus (AGNs), and their classification is crucial in astronomical research. However, due to the particularity of their samples, even a small amount of pollution can significantly increase the difficulty of quasar candidate discovery. Several studies have been conducted to classify quasars using AI techniques. \citet{gao2008support} studied the performance of SVM and KdTree in classifying stars and quasars in multiband data. \citet{richards2009eight} used Bayesian methods to classify 5546 candidate quasars in the Sloan Digital Sky Survey (SDSS). \citet{abraham2012photometric} implemented this topic using the difference boosting neural network method. \citet{jiang2013data} identified candidate stars for Catalytic Variables (CVs) from SDSS and LAMOST database spectra based on SVM and RF. \citet{schindler2017extremely} used the RF machine learning algorithm on SDSS and Wide-field Infrared Survey Explorer (WISE) photometry to classify quasar stars and estimate photometric redshift, and proposed a quasar selection algorithm and quasar candidate directory.

\paragraph{Automatic Estimation of Photometric Redshift}

\begin{figure}
	\centering
	\includegraphics[width=0.95\textwidth]{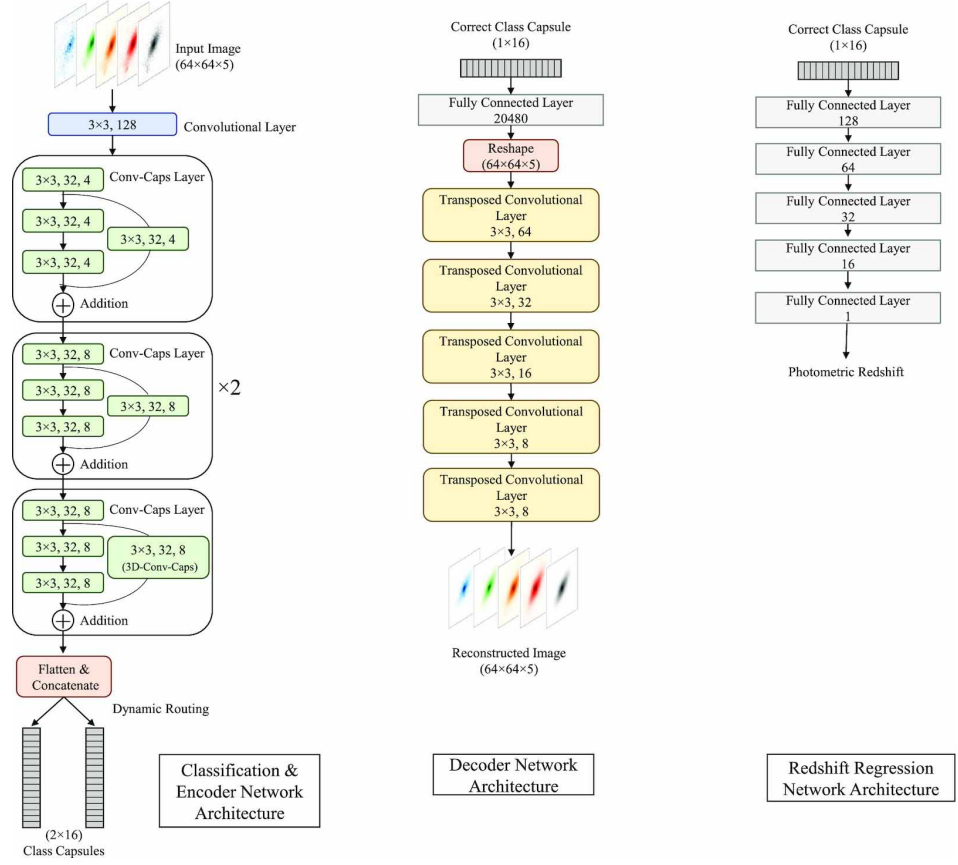}
	\caption{ A combination of three networks to jointly predict the morphology and photo-z \citep{dey2022photometric}}
	\label{fig:fig9}%
\end{figure}

Galaxy photometry redshift, the so-called photo-z, refers to the redshift of celestial objects obtained using medium and wide-band photometry or imaging data. Photometric redshifts are a key characteristic, especially for dark sources where spectral data cannot be obtained. In recent years, mounts of studies have been conducted to measure the redshift of celestial objects using AI techniques, which have shown obvious advantages in reducing cost and time consumption. The gradient-boosting tree methods like XGBoost and CATBoost \citep{humphrey2023improving,li2023photometric}, Gaussian mixture models (GMM) \citep{hatfield2020augmenting,jones2019gaussian}, KNN \citep{zhang2019new,han2021geneticknn}, SOM \citep{wilson2020photometric}, and some other supervised machine learning models \citep{bilicki2021bright,razim2021improving} have been implemented to estimate and measure photo-z from multisource data, and have achieved certain effects. Combinations of neural networks and different machine learning methods to estimate photo-z are also utilized \citep{henghes2021benchmarking,hong2023photoredshift,curran2021qso}. Meanwhile, deep learning has been a powerful strategy for assessing photo-z, for instance, based on CNN, studies \citep{dey2022photometric,zhou2022extracting,pasquet2019photometric} have realized ideal results, demonstrating the feasibility of deep learning methods for photo-z measurement and estimation, such as a combination of three networks to jointly predict the morphology and photo-z shown in Figure \ref{fig:fig9}.

\paragraph{Measurement of Stellar Parameters}

Astronomical databases require high-precision measurements to provide accurate positions, radial velocities, and physical parameters of a large number of individual stars. AI techniques have been increasingly applied in the automatic processing and measurement analysis of stellar physical parameters. Various studies have been conducted to determine stellar parameters using AI techniques. \citet{bailer1997physical} first adopted ANN methods to determine stellar atmospheric parameters based on different spectral characteristics. KNN \citep{fuentes2001prediction}, PCA and Bayesian methods \citep{bailer2011bayesian,maldonado2020hades,ciucua2021unveiling} have been used to effectively estimate stellar physical parameters. With the spurt progress of AI, different machine learning methods have been utilized to analyze stellar parameters \citep{perger2023machine,remple2021determining,passegger2022metallicities,hughes2022galah}, and some powerful pipeline tools based on machine learning have been implemented for stellar physics parameter estimation and measurement, such as ODUSSEAS \citep{antoniadis2020odusseas}, ROOSTER \citep{breton2021rooster} and SUPPNet \citep{rozanski2022suppnet}. Benefiting from a huge volume of astrophysical data, DL methods are also widely used in this field and have become a research hotspot \citep{cargile2020minesweeper,claytor2022recovery,johnson2020rotnet}, and the assessment of stellar physical parameters yields remarkable results. 

In particular, LAMOST has obtained more than 20 million spectral data, which is currently in the leading position among all telescopes in the world. AI technologies, such as machine learning and deep learning methods, are utilized to evaluate and measure stellar parameters based on mass data from LAMOST \citep{rui2019analysis,minglei2020atmospheric,zhang2020deriving,li2023estimating,bai2019machine,yang2022j,wang2020spcanet,chen2022application,li2018carbon}, and much scientific research progress has been made in the fields of searching for special celestial bodies such as lithium-rich giants, metal-poor stars, hypervelocity stars, and white dwarfs.

\section{Discussion}\label{section3}

The articles cited in this review encompass both journal papers and conference papers. The methods proposed within these works are exclusively those that have been empirically validated using either a telescope or a telescope prototype. Our review did not take into consideration methods requiring additional validation on the telescope using simulation data.

\subsection{Telescope Intelligence Research Hotspots}

AI technology has found applications in all facets of telescope operation. We categorize telescope intelligence research into six main areas: site selection, optical system calibration, observational schedule, fault diagnosis, imaging quality optimization, and database optimization. These categories are further divided into subcategories, each serving as a distinct research direction.

\begin{figure*}
	\centering
	\includegraphics[width=1\textwidth]{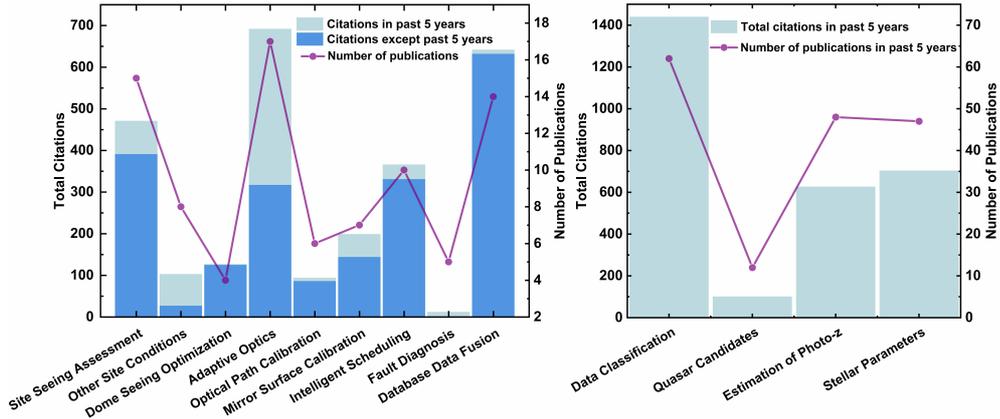}
	\caption{The right part enumerates both the quantity of publications and the accompanying citations pertaining to database data labeling, while the left part similarly enumerates the citation counts and number of publications in various other research domains. The citation count for articles published in the last five years reveals that, besides database calibration, the application of AI techniques in adaptive optical technology and site seeing assessment fields are currently research hotspots}
	\label{fig:fig10}%
\end{figure*}

To study the focal points of each research direction, we have conducted an extensive count of the journal articles published in each field, which includes the number of papers per direction, the total citation count for papers, and the citations of articles published in the past five years. The statistical results indicate that the number of articles published on database data labeling far exceeds those in other areas, making it difficult to account for them fully. We have separately listed the articles and their citation counts from the past five years in this particular field. The results are presented in the Figure \ref{fig:fig10}. The citation count for articles published in the last five years reveals that, in addition to  database calibration, the application of AI techniques in adaptive optical technology and site seeing assessment fields are currently research hotspots.

\subsection{Telescope Intelligence Research Trend}

We analyze the future research trajectory of AI technology in this domain by comparing it with traditional methods prevalent in the field, taking into account factors such as time efficiency and accuracy. We use a scoring system where AI-based methods are awarded 1 point when their performance surpasses traditional methods or accomplishes tasks that traditional methods cannot, 0 points when they match the efficacy of traditional methods, and -1 point when they underperform in comparison. The overall evaluation is obtained by summing up these two evaluation metrics. The higher the score, the more significant the advantages of AI technology become, indicating a trend towards its wider adoption in the future. A classification of various research directions based on these criteria is provided in Table \ref{tab3}. The results indicate that the use of AI technology is particularly advantageous in the optimization of dome seeing, observational planning and database data labeling.

\begin{table}[thp]
	\caption{Classification of various research directions based on proposed criteria}
	\centering
	\label{tab3}
	\setlength{\tabcolsep}{3mm}{
		\begin{tabular}{@{}cccc@{}}
			\hline
			Item & Time Cost  & Accuracy	 & Level \\
			\hline%
			Site Seeing Estimate and Prediction &   0  & 0  &    0  \\
			Assessment of Site Observation Conditions &  1  & -1  & 0  \\
			Optimization of Dome Seeing &  1  &  1  &  2  \\
			Adaptive Optics &  1  &  0  & 1 \\
			Optical Path Calibration &  1  & 0  & 1 \\
			Mirror Surface Calibration & 1 &  0 & 1 \\
			Observation Schedule & 1 &  1  & 2 \\
			Fault Diagnosis & 1  &  0 & 1 \\
			Database Data Fusion  &  1  &  0  & 1 \\
			Date Classification &  1 &  1  & 2 \\
			Preselected Quasar Candidates & 1 &  1 & 2 \\
			Photometric Infrared Evaluation & 1  &  1  & 2 \\
			Stellar Parameter Measurements &  1  &  1 & 2 \\
			\hline
	\end{tabular}}
\end{table}

\subsection{Future Hotspots of Telescope Intelligence}

As the depth of our universe exploration expands, the requirement for higher imaging sensitivity in telescopes is escalating, especially for observing more distant and dimmer celestial objects. Increasing the aperture size of telescopes can enhance their resolution, with several 30-meter class telescopes currently under construction. The technique of light interference using multiple smaller telescopes can also boost resolution while reducing the cost of constructing large-aperture telescopes.

To minimize the impact of atmospheric disturbances on telescope resolution, it is crucial to position these instruments at sites with superior seeing conditions and develop high-performance adaptive optical systems. Space telescopes, unaffected by atmospheric disturbances, can deliver imaging at the optical diffraction limit, although they are subject to the challenge of high costs. In the era of large-aperture telescopes, smaller telescopes continue to play significant roles, particularly in detecting transient sources like pulsars, gamma-ray bursts, and gravitational waves. Coordinated observations employing small lens arrays allow for uninterrupted all-sky monitoring of observational targets.

The following sections will delve into the challenges and complexities associated with ground-based large-aperture telescopes, optical interference technology, space telescopes, and small telescope arrays. These same challenges also represent the exciting frontiers for applying AI technology in the field of telescope technology in the future.

\subsubsection{Large-aperture Telescopes and Optical Interference Technology}

Currently, construction is underway on a series of 30-meter class telescopes including the Thirty Meter Telescope (TMT), the Extremely Large Telescope (ELT), and the Giant Magellan Telescope (GMT). These large-aperture telescopes offer superior light flux and angular resolution but are accompanied by more intricate structures.

Consider the ELT as an example: the optical system of the ELT comprises five mirrors. The primary mirror, spanning 39 meters in diameter, is assembled from 798 sub-mirrors, each with a 1.5-meter aperture, and requires a surface accuracy of 10 nanometers. The M4 secondary mirror, with a thickness less than 2 mm and a diameter of 2.4 meters, features 5000 magnets on the rear surface. This intricate arrangement is designed to achieve surface changes with an accuracy of 10 nanometers a thousand times per second \citep{hippler2019adaptive}. 

For 30-meter class optical telescopes, testing of large-aperture sub-mirrors presents significant challenges due to airflows that cause image blurring and difficulties in assessing surface accuracy. Active optics demand control over a greater number of mirror surfaces and actuators, hence the requirements for precision in active optical detection and control are elevated. The substantial weight of these large telescopes poses further challenges for support structures, while the increased size of the adaptive optics deformable mirrors necessitates more sophisticated manufacturing and control processes, all while maintaining high efficiency.

Optical interference technology, already deployed in telescopes such as the VLT, CHARA, KECK, and LBT, will retain its high-resolution edge, even in the era of 30-meter telescopes. The MRO Interferometer, presently under construction and employing 10 telescopes for interference imaging, is projected to achieve a resolution 100 times greater than that of the Hubble Space Telescope \citep{buscher2013conceptual}. Concurrently, the Nanjing Institute of Astronomical Optics \& Technology of the Chinese Academy of Sciences is developing an optical interference project, that incorporates three 600 mm aperture telescopes and a baseline length of 100 meters. Optical interference, due to its advantages of lengthy baselines and a wide spectrum of observable wavebands, is attracting widespread interest. However, current optical interference experiments have resolutions far below the theoretical limit, owing to factors such as detector noise and telescope vibrations \citep{eisenhauer2023advances}. As a result, the development of AI techniques to enhance the wavelength and baseline dynamic range of optical interference will be a critical area for future exploration.

\subsubsection{Space Telescope}

Space telescopes, which operate free from atmospheric disturbances, include the recently successfully launched James Webb Space Telescope (JWST), along with the planned Chinese Space Station Telescope (CSST), the Space Infrared Telescope for Cosmology and Astrophysics (SPICA), and the Large Ultraviolet Optical Infrared Surveyor (LUVOIR). However, these instruments pose unique challenges: their maintenance costs are substantial, their deployment presents significant hurdles, and they require advanced levels of automated control.

Among the key difficulties faced by space telescopes is the necessity for active optical technologies to maintain the integrity of their surfaces. This is akin to the active optical technologies employed in ground-based telescopes. Another challenge is the need to minimize the equipment size. An example of such innovation is the JWST's design, which uses low-temperature programmable slit masks for multiobject spectroscopy \citep{boker2022near}. The application of AI to replace intricate hardware devices will be a crucial area of research in the future. In addition, the high-resolution imaging generated by these telescopes produces vast amounts of data. Thus, achieving low-power, high-speed, long-distance data transmission will also be a significant focus of future research.

\subsubsection{Small Aperture Telescope Array}

Contrary to possible anticipations, the advent and construction of larger telescopes have not rendered smaller telescopes obsolete. Instead, these compact instruments have found wide-ranging applications in diverse fields, including the study of gamma-ray bursts, the detection of exoplanets, and the investigation of microlensing phenomena. By assembling arrays of small telescopes across multiple continents, researchers can perform continuous monitoring of specific astronomical targets, or alternatively, utilize wide-field small telescope arrays for sky survey observations.

Projects currently under development, such as the Panoramic Survey Telescope and Rapid Response System (Pan-STARRS), intend to facilitate rapid sky surveys by utilizing arrays of four small telescopes \citep{magnier2020pan}. Similarly, the SiTian project employs an array of small telescopes for all-sky survey observations, utilizing a network of 54 telescopes, each with a diameter of one meter \citep{chen2022optical}. This methodology is expected to catalyze significant breakthroughs in the field of time-domain astronomy. Moreover, the Stellar Observations Network Group (SONG) project is planning to construct a globally interconnected observational system by installing eight 1-meter telescopes, each with varying apertures, at fixed latitudes and longitudes in both the Northern and Southern Hemispheres \citep{grundahl2013stellar}.

The hotspot for future research in this domain will likely encompass the coordination and control of multiple telescopes for collaborative observations, the determination of optimal observational strategies, and the implementation of effective cluster control systems.

\subsubsection{The Challenge of Satellite Megaconstellations}

Ground-based telescopes are increasingly contending with the interference resulting from sunlight reflected off artificial satellites. In recent years, a multitude of satellite launch projects, including Starlink2, Kuiper, and WorldVu, have proposed ambitious plans to deploy approximately 60,000 low-Earth-orbit satellites by the year 2030 \citep{halferty2022photometric}.

Even after the implementation of sunshades, the brightness of Starlink's VisorSat version is expected to reach the 6th magnitude, causing significant disruptions to ground-based telescopes, particularly those engaged in sky survey operations \citep{hainaut2020impact}. Moreover, the Hubble Space Telescope has reported impacts stemming from satellite-reflected light \citep{kruk2023impact}. Therefore, strategies for mitigating the influence of satellites on astronomical observations will be a critical focal point for future research in the field.

\subsubsection{Large Language Models Improve Telescope Intelligence}

Large language models (LLMs), such as BERT \citep{devlin2018bert}, Llama2 \citep{touvron2023llama}, GPT-4 \citep{bubeck2023sparks}, are representative AI technologies that have attracted the most attention recently. Meanwhile, LLMs have been widely utilized in many fields such as text reading \citep{beltagy2019scibert}, medicine \citep{thirunavukarasu2023large,mesko2023imperative}, and education \citep{kasneci2023chatgpt}, demonstrating outstanding application capabilities.

Future LLMs may be able to more directly serve telescope intelligence, bringing huge potential in telescope equipment status monitoring, optimizing observation plans, etc. Additionally, the vast amount of data used in training LLMs enables them to recognize and extract meaningful information from complex astronomical data. This capability offers new avenues for accelerating scientific discovery and deepening our understanding of the universe.

\section{Conclusion}\label{section4}

This article delves into the role of AI in various aspects of telescope operation and research. These aspects include selecting telescope sites, calibrating optical systems, diagnosing faults, optimizing image quality, making observational decisions, and enhancing the intelligence of databases. The piece presents both the current focus areas and specific topics of research within the realm of telescope intelligence.

The paper provides a comprehensive statistical analysis of recent research trends. It reveals that the labeling of astronomical data within intelligent databases has become a significant hotspot. This particular field has seen the most prolific publication of papers. Additionally, there is an extensive body of published work dedicated to adaptive optical technology and site seeing assessment. Through a comparison of the time efficiency and precision of AI technology with traditional methods, the findings indicate that the use of AI technology is particularly advantageous in optimizing dome seeing, observational planning, and database data labeling.

The article concludes by projecting future advancements in telescopes. These include the development of large-aperture telescopes, optical interference technology, arrays of small telescopes, space telescopes and large language models customized for astronomy. Additionally, it addresses potential threats posed by satellite megaconstellations to telescopes. Given the evolving landscape of telescope technology, it also identifies likely areas of focus for future research.

\section*{Acknowledgements}

Thanks are given to the reviewer for the constructive comments and helpful suggestions. This work is supported by the National Nature Science Foundation of China (Grant No U1931207, 12203079, 12103072, 11973065 and 12073047), the Natural Science Foundation of Jiangsu Province (Grants No BK20221156 and BK20210988), and the Jiangsu Funding Program for Excellent Postdoctoral Talent (Grant No 2022ZB448).

\bibliography{sn-bibliography.bib}

\end{document}